 \def\CH{\hbox{{$\cal H$}}}

\def\CR{\hbox{{$\cal R$}}}

\def\eqn#1#2{\begin{equation}#2\label{#1}\end{equation}}

\def\proof{\goodbreak\noindent{\bf Proof\quad}}
\def\lform{\hbox{$\sqcup$}\llap{\hbox{$\sqcap$}}}
\def\endproof{{\ $\lform$}}

\newcommand{\Ad}{\mbox{$A^{\dagger }$}}
\newcommand{\Ads}[1]{\mbox{$A^{\dagger }_{#1}$}}
\newcommand{\Adb}{\mbox{$\bar{A}^{\dagger }$}}
\newcommand{\Adbs}[1]{\mbox{$\bar{A}^{\dagger }_{#1}$}}
\newcommand{\nuqh}{\mbox{$U_{q}(h)^{\otimes n}$}}
\newcommand{\nb}{\mbox{$BU_{q}(h)^{\underline{\otimes } n}$}}
\newcommand{\Xd}{\mbox{$\bar{X}^{\dagger }$}}
\newcommand{\Xds}[1]{\mbox{$\bar{X}^{\dagger }_{#1}$}}

\newcommand{\Xs}[1]{\mbox{$\bar{X}_{#1}$}}
\newcommand{\Ab}{\mbox{$\bar{A}$}}
\newcommand{\Abs}[1]{\mbox{$\bar{A}_{#1}$}}
\newcommand{\Hb}{\mbox{$\bar{H}$}}
\newcommand{\Hbs}[1]{\mbox{$\bar{H}_{#1}$}}
\newcommand{\Nb}{\mbox{$\bar{N}$}}
\newcommand{\Nbs}[1]{\mbox{$\bar{N}_{#1}$}}

\newcommand{\Mbs}[1]{\mbox{$\bar{M}_{#1}$}}
\newcommand{\eh}[2]{\mbox{$e^{\frac{\omega }{#2}\hbar_{#1}}$}}
\newcommand{\ehn}[2]{\mbox{$e^{-\frac{\omega }{#2}\hbar_{#1}}$}}
\newcommand{\eH}[2]{\mbox{$e^{\frac{\omega }{#2}\bar{H}_{#1}}$}}
\newcommand{\eHn}[2]{\mbox{$e^{-\frac{\omega }{#2}\bar{H}_{#1}}$}}
\newcommand{\esumh}[3]{\mbox{$e^{\frac{\omega }{#3}(\hbar_{#1} + \cdots +
\hbar_{#2})}$}}
\newcommand{\esumhn}[3]{\mbox{$e^{-\frac{\omega }{#3}(\hbar_{#1} + \cdots +
\hbar_{#2})}$}}

\newcommand{\esumH}[3]{\mbox{$e^{\frac{\omega }{#3}(\bar{H}_{#1} + \cdots +
\bar{H}_{#2})}$}}
\newcommand{\esumHn}[3]{\mbox{$e^{-\frac{\omega }{#3}(\bar{H}_{#1} + \cdots +
\bar{H}_{#2})}$}}
\newcommand{\esumHnnb}[3]{\mbox{$e^{-\frac{\omega }{#3}(H_{#1} + \cdots +
H_{#2})}$}}
\newcommand{\esumHnd}[2]{\mbox{$e^{\omega (\bar{H}_{#1} + \cdots +
\bar{H}_{#2})}$}}
\newcommand{\esumHnnd}[2]{\mbox{$e^{-\omega (\bar{H}_{#1} + \cdots +
\bar{H}_{#2})}$}}

\newcommand{\ersumHn}[3]{\mbox{$e^{-\frac{\omega }{#3}(\bar{H}_{#1} +
\bar{H}_{#2})}$}}
\newcommand{\half}[1]{\mbox{$[#1]^{\frac{1}{2}}$}}
\newcommand{\utens}{\mbox{$\underline{\otimes }$}}
\newcommand{\uS}{\mbox{$\underline{S}$}}

\documentstyle[11pt]{article}
\newtheorem{lemma}{Lemma}[section]
\newtheorem{propos}[lemma]{Proposition}

\newtheorem{theorem}[lemma]{Theorem}
\newtheorem{cor}[lemma]{Corollary}

\begin{document}\baselineskip 25pt

{\ }\hskip 4.7in DAMTP/92-39 %put here preprint number
\vspace{.5in}

\begin{center} {\bf THE BRAIDED HEISENBERG GROUP}
\baselineskip 13pt{\ }\\
{\ }\\ W.K. Baskerville\footnote{Work supported by Packer Australia Scholarship
(Cambridge Commonwealth Trust), and Overseas Research Studentship.} and
S. Majid\footnote{SERC Fellow and Drapers Fellow of Pembroke College,
Cambridge.}\\ {\ }\\
Department of Applied Mathematics\\
\& Theoretical Physics\\ University of Cambridge\\ Cambridge CB3 9EW, U.K.
\end{center}

\begin{center}
August 1992\end{center}
\vspace{10pt}
\begin{quote}\baselineskip 13pt
\noindent{\bf ABSTRACT}
We compute the braided groups and braided matrices $B(R)$ for the solution $R$
of the Yang-Baxter equation associated to the quantum Heisenberg group. We also
show that a particular extension of the quantum Heisenberg group is dual to the
Heisenberg universal enveloping algebra $U_{q}(h)$, and use this result to
derive an action of $U_{q}(h)$ on the braided groups. We then demonstrate the
various covariance properties using the braided Heisenberg group as an explicit
example. In addition, the braided Heisenberg group is found to be self-dual.
Finally, we discuss a physical application to a system of n braided harmonic
oscillators. An isomorphism is found between the n-fold braided and
unbraided tensor products, and the usual `free' time evolution is shown to be
equivalent to an action of a primitive generator of $U_{q}(h)$ on the braided
tensor product.
\end{quote}
\baselineskip 25pt

\section{Introduction}
\label{intro}

Braided groups \cite{MajalgQFT,Majex,Majskyl} are a variant of quantum groups
or superquantum groups in which the $\pm 1$ of super statistics is replaced by
a more general braiding, $\Psi $\@. This braiding replaces the ordinary twist
map, $\tau $, in the Hopf algebra (or bialgebra) axioms. This has its most
obvious effect in the tensor product algebra structure, where the usual
definition of the product (for quantum groups)
\eqn{qtp}{(a \otimes b)(c \otimes d) = a \tau (b \otimes c) d = ac \otimes bd}
must now be rewritten to incorporate the braid statistics
\eqn{btp}{(a \otimes b)(c \otimes d) = a \Psi (b \otimes c) d }
Note that the braided transposition $\Psi $ may produce a linear combination of
terms in the tensor product algebra: this is more general than the modified
transposition of superquantum groups, which differs from $\tau $ only by a
variable sign.

One of the main motivations for introducing braided groups is that they can be
used as a tool for doing quantum group calculations in a fully covariant
way \cite{Majex}. For a quantum group, it is possible to define an action of
the universal enveloping algebra on itself or the function algebra, which will
preserve {\em either\/} the product or the coproduct structure, but not both.
Braided groups have been modified in precisely such a way as to allow the full
Hopf algebra structure to be conserved by the action. The braiding arises as a
natural result of this process. All quantum groups have braided
group versions \cite{Majex}.

Here we study in detail one example of these braided groups: namely, the
braided
version of the quantum Heisenberg group. This has been chosen for study as it
provides a particularly simple example with which to demonstrate the above
properties. It is also of great physical interest, since the universal
enveloping algebra $U_{q}(h)$ is basically the `quantum group version' of the
ordinary quantum harmonic oscillator \cite{cel,GS}. The algebra
of $U_{q}(h)$ differs from the usual harmonic oscillator algebra only by the
fact that the number operator, $N = a^{\dagger}a\/$, is viewed as a primitive
generator, and the role of Planck's constant is taken over by a central,
grouplike operator (see Section~\ref{sec-mod} for details.) It is hoped that
the
close link between the Heisenberg group and such a well-known system may allow
some insight to be gained into the physical meaning of some of the mathematical
properties: in particular the braiding itself.

We begin in Section~\ref{sec-quantum} by reviewing the explicit structure of
$H_{q}(1)$, the quantum Heisenberg group of function algebra type
\cite{cel}.
We also study two extensions of it which will be useful in later sections, and
show how they arise by taking successive quotients of some general $3 \times 3$
matrices of Heisenberg type. (The two extensions are, specifically, the upper
triangular matrices of Heisenberg type $T_{+}(h)$, and a central extension
$\widetilde{H_{q}(1)}$.) In Section~\ref{sec-braided} we go on to compute the
corresponding braided groups of function algebra type. The general $3 \times 3$
braided matrices (of Heisenberg type) are calculated first, and then various
quotients $BT_{+}(h)$, $\widetilde{BH_{q}(1)}$, $BH_{q}(1)$ are taken, in
close analogy to the procedure followed in Section~\ref{sec-quantum}.

In Section~\ref{sec-mod} the covariance properties of both the quantum and
braided groups under the action of $U_{q}(h)$ are discussed. In the process,
the duality between $U_{q}(h)$ and $H_{q}(1)$ is clarified - it will be
shown that it is in fact the central extension $\widetilde{H_{q}(1)}$ which is
dual to $U_{q}(h)$, not $H_{q}(1)$ itself. Correspondingly, in the braided
case,
the centrally extended $\widetilde{BH_{q}(1)}$ is dual to the braided
universal enveloping algebra $BU_{q}(h)$. In fact, $\widetilde{BH_{q}(1)}$ is
shown to be self-dual, due to the existence of an isomorphism:
$\widetilde{BH_{q}(1)} \cong BU_{q}(h)$.

We conclude in Section~\ref{sec-app} with a physical application to an n-body
system of braided harmonic oscillators. We first look for a Fock space
representation of the n-fold braided tensor product \cite{MajalgQFT,Majex}, and
in doing so discover an isomorphism between the braided and unbraided tensor
product algebras (denoted \nb\ and \nuqh\ respectively). The characteristic
feature of the braided tensor product is its covariance under the action of
$U_{q}(h)$. We use this action to define a time-evolution of the (braided)
system. The isomorphism is then used to map the $U_{q}(h)$-action and resulting
time-evolution across to the unbraided tensor algebra $U_{q}(h)^{\otimes n}$,
where it is found to correspond to that induced by the free Hamiltonian. Thus
it
is shown that the usual time evolution of n uncoupled harmonic oscillators, is
equivalent to the action of a primitive generator of $U_{q}(h)$ on the braided
harmonic oscillators.

\section{The Quantum Heisenberg Group}
\label{sec-quantum}

We begin by recalling the general FRT construction \cite{FRT} for quantum
groups
of function algebra type. To any matrix
$R \in M_{n}({\bf C}) \otimes M_n({\bf C})$ which is a solution of the Quantum
Yang-Baxter Equation (QYBE), there can be associated a non-commutative algebra
generated by 1 and n$^{2}$ generators $\{t^{i}\,_{j}\}$, modulo certain
relations. When coupled with a standard definition of the coproduct and counit
structure, this then forms the quantum (semi)group $A(R)$ associated with
$R$\@.
If $T$ is the ${\rm n \times n}$ matrix of the generators $\{t^{i}\,_{j}\}$,
then the defining relations can be written as follows in matrix form
\eqn{FRTdef}{R T_{1} T_{2} = T_{2} T_{1} R}
\eqn{coprod}{\Delta(t^{i}\,_{j}) = \sum_{k} t^{i}\,_{k} \otimes t^{k}\,_{j},\ \
\ \ \ \ \ \ \varepsilon (t^{i}\,_{j}) = \delta^{i}\,_{j} }
where $T_{1} = T \otimes {\bf 1}$ and $T_{2} = {\bf 1} \otimes T$\@.

We use the explicit form of the Heisenberg $R$-matrix given by
Celeghini et al. \cite{cel}
\eqn{Rmat}
{R = \left( \begin{array}{c|c|c}
		  I_{3} & \begin{array}{ccc}
			      0 & 0 & 0 \\
			      0 & 0 & \omega \\
			      0 & 0 & 0
			  \end{array} & \begin{array}{ccc}
					    0 & 0 & 0 \\
					    0 & -\frac{\omega }{2} & 0 \\
					    0 & 0 & 0
					\end{array} \\
		  \hline
		  0 & \begin{array}{ccc}
			  1 & 0 & -\frac{\omega }{2} \\
			  0 & 1 & 0 \\
			  0 & 0 & 1
		      \end{array} & 0 \\
		  \hline
		  0 & 0 & I_{3}
	      \end{array} \right) }
The most general quantum group associated with this $R$ is obtained by taking
$T$ in Equation~(\ref{FRTdef}) to be a general $3 \times 3$ matrix (ie. with 9
indeterminates $\{t^{i}\,_{j}\}$). However, we can also consider different
quotients of this general group by imposing additional (consistent) relations
on
the generators $\{t^{i}\,_{j}\}$. The most general quotient we consider
explicitly involves what naturally may be called the upper triangular matrices
of Heisenberg type. We denote it $T_{+}(h)$. This is obtained by setting
\eqn{utquo}{t^{i}\,_{j} = 0,\ \ \ \ \ \ \ \ i<j}
This can be done while maintaining consistency with the FRT construction, since
the generators which are being set to zero generate a bi-ideal of $A(R)$\@. To
see this, consider the general form of the coproduct:
$\Delta(t^{i}\,_{j}) = t^{i}\,_{k} \otimes t^{k}\,_{j}$. It is evident that if
$i<j$, then either $i<k$ or $k<j$, for all $k$. So span$\{t^{i}\,_{j} :i<j\}$
will always define a co-ideal of $A(R)$ (for any $R$\@.) The upper triangular
nature of the Heisenberg $R$-matrix ensures that setting these generators
to zero does not make the relations too trivial.

Using the following notation for remaining matrix of generators
\eqn{Tform}{T = \left( \begin{array}{ccc}
 				d_{1} & \alpha & \beta \\
				0 & \gamma & \delta \\
				0 & 0 &d_{2}
			\end{array} \right) }
the algebra relations can be written
\eqn{dequal}{d_{1} = d_{2}\: (= d)}
\eqn{utalgrel}{\alpha \beta - \beta \alpha = \frac{\omega }{2} \alpha d,
\ \ \ \ \ \ \ \ \alpha \delta - \delta \alpha = 0,
\ \ \ \ \ \ \ \ \beta \delta - \delta \beta = -\frac{\omega }{2} \delta d }
\eqn{utgamrel}{\alpha \gamma - \gamma \alpha = \beta \gamma - \gamma \beta =
\delta \gamma - \gamma \delta = d \gamma - \gamma d = 0 }
\eqn{utdrel}{\alpha d - d \alpha = \beta d - d \beta = \delta d - d \delta = 0}
Note that since both $d$ and $\gamma $ are grouplike, and commute with all
other
generators, the determinant $D = dd\gamma $ is also central and grouplike, as
is
its inverse $D^{-1}$. It is therefore possible to formally adjoin $D$ and
$D^{-1}$ to the algebra, with the relations $D D^{-1} = 1 = D^{-1} D$\@. This
then allows the antipode stucture to be defined
\eqn{utSone}{S(\alpha ) = -D^{-1} d \alpha ,\ \ \ \ \ \ \ \
S(\beta ) = -D^{-1} \gamma \beta + D^{-1} \alpha \delta ,\ \ \ \ \ \ \ \
S(\delta ) = -D^{-1} d \delta }
\eqn{utStwo}{S(\gamma ) = D^{-1} d d,\ \ \ \ \ \ \ \
S(d) = D^{-1} d \gamma ,}
thus giving $T_{+}(h)$ the full Hopf algebra structure. Note that the FRT
construction guarantees that $A(R)$ will always be a bialgebra, but the
existence of an antipode is not necessarily assured.

A further quotient can be taken by setting the diagonal elements of
$T$ to 1.
\eqn{heisquo}{t^{i}\,_{i} =1}
Note that this can be done while retaining the Hopf algebra structure, since
$\gamma $ and $d$ are both central and grouplike, and can therefore be set to 1
without invalidating any of the axioms. The algebra and antipode relations then
reduce to
\eqn{heisalgrel}{\alpha \beta - \beta \alpha = \frac{\omega }{2} \alpha,
\ \ \ \ \ \ \ \ \alpha \delta - \delta \alpha = 0,\ \ \ \ \ \ \ \
\beta \delta - \delta \beta = -\frac{\omega}{2} \delta }
\eqn{heisS}{S(\alpha ) = -\alpha ,\ \ \ \ \ \ \ \
S(\beta ) = -\beta = \alpha \delta ,\ \ \ \ \ \ \ \ S(\delta ) = -\delta }
It is this quotient which is commonly known \cite{cel} as the quantum
Heisenberg
group, $H_{q}(1)$.

Valid Hopf algebras could also be obtained by taking quotients intermediate
between $T_{+}(h)$ and $H_{q}(1)$; ie.\ by setting some, but not all, of the
diagonal elements to 1. Of particular interest to us is the case where $d$ is
set to 1, but $\gamma $ is retained. As we shall see in a later section, this
algebra turns out to be dual to the enveloping Heisenberg algebra. We
shall call it the extended quantum Heisenberg group, and denote it
$\widetilde{H_{q}(1)}$. Explicitly it looks like
\eqn{extheisalgrel}{\alpha \beta - \beta \alpha = \frac{\omega}{2} \alpha ,
\ \ \ \ \ \ \ \ \alpha \delta - \delta \alpha = 0,\ \ \ \ \ \ \ \
\beta \delta - \delta \beta = -\frac{\omega}{2} \delta }
\eqn{extheisgamrel}{\alpha \gamma - \gamma \alpha = \beta \gamma - \gamma \beta
 = \delta \gamma - \gamma \delta = 0 }
\eqn{extheisS}{S(\alpha ) = -\gamma ^{-1} \alpha ,\ \ \ \ \ \ \ \
S(\beta ) = -\beta + \gamma ^{-1} \alpha \delta ,\ \ \ \ \ \ \ \
S(\delta ) = -\gamma ^{-1} \delta ,\ \ \ \ \ \ \ \
S(\gamma ) = \gamma ^{-1} }
where $\gamma ^{-1}$, with relations $\gamma \gamma ^{-1} = 1 = \gamma ^{-1}
\gamma $, can be formally adjoined to the algebra in a similar way to $D$ and
$D^{-1}$ for $T_{+}(h)$, to allow convenient expression of the antipode.

\section{The Braided Heisenberg group}
\label{sec-braided}

The general construction for $B(R)$ \cite{Majex}, the braided matrices (or,
after quotienting, the braided group) associated with $R$, is similar to that
for $A(R)$. It consists again of $n^{2}$ generators ${u^{i}\,_{j}}$ constrained
by matrix relations. The definition of the coproduct and counit structure is
also again of matrix type
\eqn{brcoco}{\underline{\Delta} u^{i}\,_{j} = u^{i}\,_{k} \otimes u^{k}\,_{j},
\ \ \ \ \ \ \ \ \underline{\varepsilon} (u^{i}\,_{j}) = \delta^{i}\,_{j} }
As was pointed out in the Introduction, the tensor product algebra structure is
quite different to that for quantum groups, due to the inclusion of the braided
transposition. This necessitates a modification of the product on the
generators; that is, $B(R)$ will in general be different as an algebra to the
corresponding $A(R)$\@. The defining algebra relations and braiding are as
given
in \cite{Majex,Majskyl} in matrix form, using the same notation as in
Section \ref{sec-quantum}.
\eqn{braidFRT}{R_{21} U_{1} R_{12} U_{2} = U_{2} R_{21} U_{1} R_{12} }
\eqn{braidingdef}{\Psi (u^{I} \otimes u^{K} ) = u^{L} \otimes u^{J}
\Psi ^{K}\,_{L}\,^{I}\,_{J} }
where $I$, $J$, etc.\ are multi-indices $(i_{0} ,i_{1} )$ for the components
$u^{i_{0} }\,_{i_{1} }$, $R_{12} = R$, $R_{21} = \tau (R_{12})$.
Equation~(\ref{braidFRT}) is also known in various other contexts.
$\Psi ^{K}\,_{L}\,^{I}\,_{J}$ is defined as follows in terms of $R$
\eqn{psi}{\Psi
^{(i_{0},i_{1})}\,_{(j_{0},j_{1})}\,^{(k_{0},k_{1})}\,_{(l_{0},l_{1})} =
R^{k_{0}}\,_{a}\,^{d}\,_{j_{0}} R^{-1}\,^{a}\,_{l_{0}}\,^{j_{1}}\,_{b}
R^{l_{1}}\,_{c}\,^{b}\,_{i_{1}} \tilde{R}^{c}\,_{k_{1}}\,^{i_{0}}\,_{d} }
where $\tilde{R} = ((R^{t_{2}})^{-1})^{t_{2}}$, with $t_{2}$ denoting transpose
in the second factor.

We compute the braided matrices of Heisenberg type, $BM_{h}(3)$, defined as
$B(R)$ for the
Heisenberg $R$-matrix shown in Equation (\ref{Rmat}). The computer program
REDUCE was used for this calculation. We find that considerable
symmetry is apparent in the algebra relations thus obtained, even for these
general matrices before any quotients are taken, and it may therefore be of
interest to record them in full. Firstly, it is found that $u^{3}\,_{1}$
commutes with all other generators
\eqn{u31comm}{u^{3}\,_{1} u^{i}\,_{j} - u^{i}\,_{j} u^{3}\,_{1} = 0}
The diagonal elements and $u^{1}\,_{3}$ commute among themselves
\eqn{diagcomm}{u^{i}\,_{i} u^{j}\,_{j} - u^{j}\,_{j} u^{i}\,_{i} = 0}
\eqn{u13comm}{u^{i}\,_{i} u^{1}\,_{3} - u^{1}\,_{3} u^{i}\,_{i} = 0}
and a certain symmetry can be observed in their commutators with the remaining
generators
\eqn{diagothcomm}{u^{i}\,_{i} u^{j}\,_{k} - u^{j}\,_{k} u^{i}\,_{i} =
(-1)^{i+k}\, c_{i} \,\omega \,u^{3}\,_{1} u^{j}\,_{k}\ \ \ \ \ \ \ \
c_{i} = \left\{ \begin{array}{ll}
			\frac{1}{2} & i = 1,3 \\
			1 & i = 2
		 \end{array} \right. }
\eqn{u13othcomm}{u^{1}\,_{3} u^{j}\,_{k} - u^{j}\,_{k} u^{1}\,_{3} =
\frac{\omega ^{2}}{4}\, u^{3}\,_{1} u^{j}\,_{k} + (-1)^{j}\, u^{j}\,_{k}\,
( u^{1}\,_{1} + u^{3}\,_{3} )}
where $u^{j}\,_{k} = u^{1}\,_{2}, u^{2}\,_{1}, u^{2}\,_{3}, u^{3}\,_{2}$. The
relations of these generators among themselves are
\eqn{othcommone}{u^{1}\,_{2} u^{3}\,_{2} - u^{3}\,_{2} u^{1}\,_{2} =
-\frac{\omega }{2}\, (u^{3}\,_{2})^{2} }
\eqn{ot1}{u^{2}\,_{1} u^{2}\,_{3} - u^{2}\,_{1} u^{2}\,_{3} =
-\frac{\omega }{2}\, (u^{2}\,_{1})^{2} }
\eqn{ot2}{u^{2}\,_{1} u^{2}\,_{3} - u^{2}\,_{3} u^{2}\,_{1} =
\omega \,(u^{3}\,_{1})^{2} }
\eqn{othcommtwo}{u^{1}\,_{2} u^{2}\,_{1} - u^{2}\,_{1} u^{1}\,_{2} =
\frac{\omega }{2}\, u^{3}\,_{1} ( u^{2}\,_{2} - 2 u^{1}\,_{1} ) }
\eqn{ut3}{u^{3}\,_{2} u^{2}\,_{3} - u^{2}\,_{3} u^{3}\,_{2} =
\frac{\omega }{2}\, u^{3}\,_{1} (u^{2}\,_{2} - 2 u^{3}\,_{3})}
\eqn{othcommthree}{u^{1}\,_{2} u^{2}\,_{3} - u^{2}\,_{3} u^{1}\,_{2} =
-\frac{\omega^{2}}{2}\, u^{3}\,_{1} (u^{1}\,_{1} - u^{2}\,_{2}) +
\omega \,u^{1}\,_{1} (u^{2}\,_{2} - u^{3}\,_{3}) +
\frac{\omega }{2} \,(u^{2}\,_{3} u^{3}\,_{2} - u^{1}\,_{2} u^{2}\,_{1})}
Also, some additional relations are imposed by the fact that the defining
matrix
equation gives two different values for some commutators, which must be set
equal
\eqn{vanrel}{u^{3}\,_{1} u^{1}\,_{2} = u^{1}\,_{1} u^{3}\,_{2},\ \ \ \ \ \ \ \
u^{3}\,_{1} u^{2}\,_{2} = u^{2}\,_{1} u^{3}\,_{2},\ \ \ \ \ \ \ \
u^{3}\,_{1} u^{2}\,_{3} = u^{2}\,_{1} u^{3}\,_{3} }
\eqn{nonvanrel}{u^{2}\,_{3} u^{3}\,_{2} - u^{2}\,_{1} u^{1}\,_{2} =
u^{2}\,_{2} (u^{3}\,_{3} - u^{1}\,_{1}) }

\begin{lemma}
\label{lem-uppt}
The sub-diagonal elements of $U$, ie.\ the generators $u^{2}\,_{1}$,
$u^{3}\,_{1}$ and $u^{3}\,_{2}$, can be set to zero in a manner consistent with
both the algebra relations and the braiding.
\end{lemma}
\proof
We first note that consistency with the algebra and braiding is all that needs
to be checked, since, as was pointed out in Section~\ref{sec-quantum}, setting
below-diagonal elements to zero will always be consistent with a coproduct
structure of matrix type. It can be seen from the algebra relations that all
commutators involving any of these three generators have right hand sides which
also go to zero in this quotient. Of the four remaining relations, three are
empty, and the other gives the condition
\eqn{zeroreq}{u^{2}\,_{2} (u^{3}\,_{3} - u^{1}\,_{1}) = 0}
If this condition is satisfied, then the quotient can be seen to be consistent
with the algebra relations. Since it is obviously desirable to retain
$u^{2}\,_{2}$ in the algebra, we shall set
\eqn{defnd}{u^{1}\,_{1} = u^{3}\,_{3} = d}
Consistency with the braiding must now be checked. Since there are 81 braid
relations, we will not show them explicitly, but merely note that all
braided transpositions of tensor products involving $u^{2}\,_{1}$,$u^{3}\,_{1}$
or $u^{3}\,_{2}$ contain at least one of these generators in all terms on the
right hand side of the equation. So, having imposed the
condition~(\ref{defnd}), no inconsistencies arise on taking the desired
quotient.
\endproof

We shall call the quotient described by Lemma~(\ref{lem-uppt}) the `braided
upper triangular matrices of Heisenberg type', and denote it $BT_{+}(h)$. To
show its structure explicitly, it will be convenient to adopt the same notation
as was defined in Equation~(\ref{Tform}) for the quantum group $T_{+}(h)$.
The first thing to note, then, is that the diagonal elements $d$ and $\gamma $
are bosonic generators in the sense that
\eqn{diagbos}{\Psi (f \otimes \gamma ) = \gamma \otimes f,\ \ \ \ \ \ \ \
\Psi (\gamma \otimes f ) = f \otimes \gamma,\ \ \ \ \ \ \ \
\Psi (f \otimes d ) = d \otimes f,\ \ \ \ \ \ \ \
\Psi (d \otimes f ) = f \otimes d\ \ \ \ \ \ \ \ \forall f.}
These elements are also central
\eqn{diagcen}{\gamma f = f \gamma,\ \ \ \ \ \ \ \ df = fd \ \ \ \ \ \ \ \
\forall f.}
The other algebra relations are
\eqn{utbralgrel}{\alpha \beta - \beta \alpha = \omega d \alpha,\ \ \ \ \ \ \ \
\alpha \delta - \delta \alpha = \omega d (\gamma - d),\ \ \ \ \ \ \ \
\beta \delta - \delta \beta = \omega d \delta }
and the rest of the braiding is given by
\eqn{braiding}{
\begin{array} {llllll}
\Psi (\alpha \otimes \alpha ) & = & \alpha \otimes \alpha &
\Psi (\beta \otimes \delta) & = & \delta \otimes \beta +
\omega \,(d - \gamma ) \otimes \delta   \nonumber \\
\Psi (\alpha \otimes \beta ) & = & \beta \otimes \alpha +
\omega \,\alpha \otimes (d - \gamma ) &
\Psi (\delta \otimes \alpha ) & = & \alpha \otimes \delta   \nonumber \\
\Psi (\alpha \otimes \delta ) & = & \delta \otimes \alpha +
\omega \,(d - \gamma ) \otimes (\gamma - d) &
\Psi (\delta \otimes \beta ) & = & \beta \otimes \delta  \nonumber \\
\Psi (\beta \otimes \alpha ) & = & \alpha \otimes \beta &
\Psi (\delta \otimes \delta ) & = & \delta \otimes \delta  \nonumber \\
\Psi (\beta \otimes \beta ) & = & \beta \otimes \beta +
\omega \,\alpha \otimes \delta  & & &
\end{array} }

\begin{propos}
\label{prop-brS}
The braided matrices $BT_{+}(h)$ have a braided determinant $D = dd\gamma $
which can be adjoined to the algebra, as can also its inverse $D^{-1}$\@. The
braided antipode on $BT_{+}(h)$ is then given by:
\[\uS(\alpha ) = -D^{-1} d \alpha ,\ \ \ \ \ \ \ \
\uS(\beta ) = -D^{-1} \gamma \beta + D^{-1} \alpha \delta ,\ \ \ \ \ \ \ \
\uS(\delta ) = -D^{-1} d \delta \]
\[\uS(\gamma ) = D^{-1} d d,\ \ \ \ \ \ \ \
\uS(d) = D^{-1} d \gamma .\]
\end{propos}
\proof
Since the elements $d$ and $\gamma$ are both central, grouplike and bosonic,
$D$ and $D^{-1}$ will also have these properties. They can therefore be
formally
adjoined to the algebra, with relations $D D^{-1} = 1 = D^{-1} D$\@. That the
antipode is as given by Proposition~(\ref{prop-brS}) can be verified by direct
substitution into the defining axiom
\eqn{Sdefeqn}{.(\uS \otimes id) \circ \underline{\Delta} a =
\underline{\varepsilon} a.}
\endproof

Note that it is special to the present example that the antipode on the braided
group $BT_{+}(H)$
is the same on the generators as that of the quantum group $T_{+}(h)$. Such
similarity is not generally observed between quantum groups and their braided
counterparts. The similarity does {\em not\/} however, extend to
higher order products of the generators. The antipode on $BT_{+}(h)$ is truly
braided in the sense that
\eqn{Sisbr}{\uS(ab) = .(\uS \otimes \uS) \circ \Psi(a \otimes b) }
whereas for a quantum group the antipode extends as a straightforward
anti-algebra map
\eqn{Snotbr}{S(ab) = .(S \otimes S) \circ \tau(a \otimes b) = S(b) S(a) }

{}From Equations~(\ref{diagbos})~and~(\ref{diagcen}) it is evident that the
further quotient
\eqn{brheisquo}{\gamma = d = 1}
can be taken, to obtain the braided version of the usual quantum Heisenberg
group, which we shall call $BH_{q}(1)$\@. Its algebra relations, braiding and
antipode are given by
\eqn{brheisalgrel}{\alpha \beta - \beta \alpha = \omega \alpha,\ \ \ \ \ \ \ \
\alpha \delta - \delta \alpha = 0,\ \ \ \ \ \ \ \
\beta \delta - \delta \beta = \omega \delta }
\eqn{quobraiding}{
\begin{array}{llllll}
\Psi (\alpha \otimes \alpha ) & = & \alpha \otimes \alpha &
\Psi (\beta \otimes \delta) & = & \delta \otimes \beta \nonumber \\
\Psi (\alpha \otimes \beta ) & = & \beta \otimes \alpha &
\Psi (\delta \otimes \alpha ) & = & \alpha \otimes \delta  \nonumber \\
\Psi (\alpha \otimes \delta ) & = & \delta \otimes \alpha &
\Psi (\delta \otimes \beta ) & = & \beta \otimes \delta \nonumber \\
\Psi (\beta \otimes \alpha ) & = & \alpha \otimes \beta &
\Psi (\delta \otimes \delta ) & = & \delta \otimes \delta \nonumber \\
\Psi (\beta \otimes \beta ) & = & \beta \otimes \beta +
\omega \alpha \otimes \delta  & & &
\end{array}  }
\eqn{brheisS}{\uS(\alpha ) = -\alpha ,\ \ \ \ \ \ \ \
\uS(\beta ) = -\beta = \alpha \delta ,\ \ \ \ \ \ \ \ \uS(\delta ) = -\delta }
A comparison of Equations~(\ref{brheisalgrel}), (\ref{quobraiding}) and
(\ref{brheisS}) with (\ref{heisalgrel}) and (\ref{heisS}) reveals that this
algebra is remarkably similar to that of the quantum Heisenberg group,
$H_{q}(1)$. Also the braiding, while taking an extremely simple form, remains
non-trivial in this quotient, ie\@. $\Psi^{2} \not= id$. $BH_{q}(1)$, then, is
probably the simplest example available of a non-trivial braided group.

The braided Heisenberg group can be extended in the same way as the quantum
group by retaining $\gamma $ and setting only $d$ to 1. Let us denote this
`extended Heisenberg group' by $\widetilde{BH_{q}(1)}$. It has the explicit
algebra structure
\eqn{extbrheisalgrel}{\alpha \beta - \beta \alpha = \omega \alpha,\ \ \ \ \ \ \
\ \alpha \delta - \delta \alpha = \omega (\gamma - 1),\ \ \ \ \ \ \ \
\beta \delta - \delta \beta = \omega \delta.}
The braided antipode is
\eqn{extbrheisS}{\uS(\alpha ) = -\gamma ^{-1} \alpha ,\ \ \ \ \ \ \ \
\uS(\beta ) = -\beta + \gamma ^{-1} \alpha \delta ,\ \ \ \ \ \ \ \
\uS(\delta ) = -\gamma ^{-1} \delta ,\ \ \ \ \ \ \ \
\uS(\gamma ) = \gamma ^{-1} }
where $\gamma^{-1}$ has been formally adjoined to
the algebra as a central, grouplike and bosonic element, with relations
$\gamma \gamma^{-1} = 1 = \gamma^{-1} \gamma$. Since $\gamma$ is bosonic, its
braiding with the other generators is just simple transposition, as given by
Equation~(\ref{diagbos}). The rest of the braiding can easily be obtained from
Equation~(\ref{braiding}), by setting $d$ to 1. Thus, as a bialgebra,
$\widetilde{BH_{q}(1)}$ differs from $BH_{q}(1)$ only by the addition of this
central, grouplike and bosonic element $\gamma$, which appears on the right in
Equation~(\ref{extbrheisalgrel}), and in one of the braiding relations. This
extension, however, plays an important role in the next section.

\section{Braided Heisenberg groups as $U_{q}(h)$-modules}
\label{sec-mod}

In this section we show that $BM_{h}(3)$, $BT_{+}(h)$, $\widetilde{BH_{q}(1)}$
and $BH_{q}(1)$ are all objects in the same braided category, namely, the
category of the representations of $U_{q}(h)$, the quantum enveloping
Heisenberg
algebra. This follows from the general theory in \cite{MajalgQFT,Majex}, but we
feel it is of interest to see it explicitly for the present case.

To demonstrate this we first need the explicit structure of the quantum
group $U_{q}(h)$, as given by \cite{cel,GS}
\eqn{Ualg}{[A,A^{\dagger}] = \frac{1}{\omega} (e^{\omega H/2} - e^{-\omega
H/2})
\ \ \ \ \ \ \ \ [N,A] = - A \ \ \ \ \ \ \ \
[N,A^{\dagger}] = A^{\dagger}\ \ \ \ \ \ \ \ H {\rm central} }
\begin{eqnarray} \label{Ucoprod}
\Delta A & = & e^{-\omega H/4} \otimes A + A \otimes e^{\omega H/4} \nonumber
\\
\Delta A^{\dagger} & = & e^{-\omega H/4} \otimes A^{\dagger} + A^{\dagger}
\otimes e^{\omega H/4} \nonumber \\
\Delta N & = & N \otimes 1 + 1 \otimes N \nonumber \\
\Delta H & = & H \otimes 1 + 1 \otimes H
\end{eqnarray}
\eqn{UcounitS}{\varepsilon X = 0 \ \ \ \ \ \ \ \ SX = -X \ \ \ \ \ \ \ \
X = A, A^{\dagger}, N, H.}
The quasitriangular structure \cite{Drinqg}, or universal $R$-matrix, is given
by
\eqn{algformRmat}{\CR = e^{-\omega / 2(H \otimes N + N \otimes H)}
e^{\omega (e^{\omega H/4} A \otimes e^{-\omega H/4} A^{\dagger} ) } }
The notation above is the same as that used by \cite{cel}, but we differ from
them in taking $N$ to be a primitive generator (they do note this as a
possibility), as was done by \cite{GS}.

The statement that $BM_{h}(3)$, etc.\ are all objects in the category of
representations of this quantum group means that $U_{q}(h)$ acts on all of
these
braided groups, and that this action is respected by all maps within each group
(ie.\ all the braided Hopf algebra operations ($\underline{\Delta},
\underline{.}, \underline{S}, \underline{\varepsilon}$) are covariant under
this
action.) It is worth noting that the quantum groups of function algebra type
($T_{+}(h)$, $\widetilde{H_{q}(1)}$ and $H_{q}(1)\:$) are $U_{q}(h)$-module
coalgebras in this sense, under the quantum coadjoint action. Their products,
however, do not respect this action. It is the modified product of the braided
versions which makes them into $U_{q}(h)$-module Hopf algebras, as explained in
general in \cite{Majrev}.

\begin{propos}
\label{prop-pairing}
$\widetilde{H_{q}(1)}$ is paired to $U_{q}(h)$ .
\end{propos}
\proof
To show that $U_{q}(h)$ and $\widetilde{H_{q}(1)}$ are paired, it is necessary
to check that the following pairing relations
\eqn{p1}{\langle \phi \psi \,, a \rangle =
\langle \phi \otimes \psi \,,\Delta a \rangle}
\eqn{p2}{\langle 1 \,, a \rangle = \varepsilon (a) }
\eqn{p3}{\langle \Delta \phi \,, a \otimes b \rangle = \langle \phi \,, ab
\rangle }
\eqn{p4}{\varepsilon (\phi ) = \langle \phi \,, 1 \rangle}
\eqn{p5}{\langle S\phi \,, a \rangle = \langle \phi \,, Sa \rangle}
are satisfied, where $\phi ,\psi \in U_{q}(h),\ a,b \in H_{q}(1)$. We define
the
pairing using the matrix representation of $U_{q}(h)$ given by \cite{cel}
\begin{eqnarray} \label{matrices}
\rho (A) = \left( \begin{array}{ccc}
			0 & 1 & 0 \\
			0 & 0 & 0 \\
			0 & 0 & 0
	   \end{array} \right)
&\ \ \ \ \ \ \ \ & \rho (A^{\dagger}) = \left( \begin{array}{ccc}
							0 & 0 & 0 \\
							0 & 0 & 1 \\
							0 & 0 & 0
					\end{array} \right)  \nonumber \\
 & & \nonumber \\
\rho (H) = \left( \begin{array}{ccc}
			0 & 0 & 1 \\
			0 & 0 & 0 \\
			0 & 0 & 0
		\end{array} \right)
&\ \ \ \ \ \ \ \ & \rho (N) = \left( \begin{array}{ccc}
						0 & 0 & 0 \\
						0 & 1 & 0 \\
						0 & 0 & 0
				\end{array} \right)
\end{eqnarray}
taking
\eqn{pairingdef}{\langle X \,, t^{i}\,_{j} \rangle = \rho (X)^{i}\,_{j}
\ \ \ \ \ \ \ \ X = A,A^{\dagger},H,N. }
The relation~(\ref{p2}) can then immediately be seen to hold on the generators
\eqn{p2check}{\langle 1 \,, t^{i}\,_{j} \rangle = I_{3}\,^{i}\,_{j}
= \delta ^{i}\,_{j} = \varepsilon (t^{i}\,_{j} ).}
Also,
\eqn{p4check}{\langle X \,, 1 \rangle = \langle X \,, t^{1}\,_{1} \rangle =
\langle X \,, t^{3}\,_{3} \rangle = 0 = \varepsilon X }
so that (\ref{p4}) is also seen to hold. Note that it is this relation which
would fail to be satisfied were $\gamma $ not retained, since
\eqn{gammfail}{\langle N \,, t^{2}\,_{2} \rangle = 1 \not= 0 .}
The relations (\ref{p1}) and (\ref{p3}) can be used to define the extension
of the pairing to higher order products of the generators, according to
\eqn{Ualgpair}{\langle XY \,, t^{i}\,_{j} \rangle = \rho (XY)^{i}\,_{j}
= \rho (X)^{i}\,_{k} \:\rho (X)^{k}\,_{j} }
\eqn{talgpair}{\langle X \,, t^{i}\,_{j} t^{k}\,_{l} \rangle =
\langle X_{(1)} \,, t^{i}\,_{j} \rangle \langle X_{(2)} \,, t^{k}\,_{l} \rangle
= \rho (X_{(1)})^{i}\,_{j} \:\rho (X_{(2)})^{k}\,_{l} }
where $\Delta X = X_{(1)} \otimes X_{(2)} $. The consistency of these
definitions with the algebra relations of $U_{q}(h)$ and $\widetilde{H_{q}(1)}$
respectively must then be checked. It can easily be verified that the
definition~(\ref{Ualgpair}) respects the algebra~(\ref{Ualg}) simply by
multiplying out the matrices~(\ref{matrices}). The consistency of the second
definition is most easily shown by a direct computation of all such pairings.
The relevant non-zero cases are
\begin{eqnarray} \label{nztalgpairs}
\langle A \,, \alpha \beta \rangle & = & \frac{\omega }{4}\ \ \ \ \ \ \ \
\langle A \,, \beta \alpha \rangle = -\frac{\omega }{4} \nonumber \\
\langle A^{\dagger} \,, \beta \delta \rangle & = & -\frac{\omega }{4}
\ \ \ \ \ \ \ \
\langle A^{\dagger} \,, \delta \beta \rangle = \frac{\omega }{4} \nonumber \\
\langle A \,, \alpha \gamma \rangle & = & \langle A \,, \gamma \alpha \rangle =
\langle A \,, \alpha \gamma^{-1} \rangle = \langle A \,, \gamma^{-1} \alpha
\rangle = 1 \nonumber \\
\langle A^{\dagger} \,, \delta \gamma \rangle & = & \langle A^{\dagger} \,,
\gamma \delta \rangle = \langle A^{\dagger} \,, \delta \gamma^{-1} \rangle =
\langle A^{\dagger} \,, \gamma^{-1} \delta \rangle = 1 \nonumber \\
\langle H \,, \beta \gamma \rangle & = & \langle H \,, \gamma \beta \rangle =
\langle H \,, \beta \gamma^{-1} \rangle = \langle H \,, \gamma^{-1} \beta
\rangle =1
\end{eqnarray}
Consistency with the algebra relations~(\ref{extbrheisalgrel}) is then easily
checked. To check the last pairing relation, it is first necessary to define
the pairing between the generators of $U_{q}(h)$ and $\gamma^{-1}$. This can be
done using the inverse property $\gamma \gamma^{-1} = 1 = \gamma^{-1} \gamma$,
and equation~(\ref{talgpair})
\eqn{gamminpairdef}{\langle X \,, 1 \rangle =
\langle X \,, \gamma^{-1} \gamma \rangle = \sum
\langle X _{(1)}\,, \gamma^{-1} \rangle \:\rho (X_{(2)})^{2}\,_{2} \ \ \ \ \ \
\ \
X = A,A^{\dagger},H,N,1. }
where $\Delta X = \sum X_{(1)} \otimes X_{(2)} $.
The pairings $\langle X _{(1)}, \gamma^{-1} \rangle $ are the only unknowns in
this set of equations, and are thus determined by them. They are, explicitly
\begin{eqnarray} \label{gamminpairings}
\langle A \,, \gamma^{-1} \rangle = 0 & \ \ \ \ \ \ \ \
\langle A^{\dagger} \,, \gamma^{-1} \rangle = 0 &\ \ \ \ \ \ \ \
\langle N \,, \gamma^{-1} \rangle = -1 \nonumber \\
\langle H \,, \gamma^{-1} \rangle = 0 & \ \ \ \ \ \ \ \
\langle 1 \,, \gamma^{-1} \rangle = 1 & \
\end{eqnarray}
The relation~(\ref{p5}) can then easily be checked directly, case by case, and
is found to hold. Thus it is established that $U_{q}(h)$ and
$\widetilde{H_{q}(1)}$ are paired.
\endproof

By inspection, the pairing between $U_{q}(h)$ and $\widetilde{H_{q}(1)}$ can be
seen to be non-degenerate (no generator pairs as zero with all other generators
on the other side). Note that all quantum groups of function algebra
type $BM_{h}(3)$ and $BT_{+}(h)$ are also paired to $U_{q}(h)$, although in
these cases the pairing is degenerate even at generator level. $H_{q}(1)$ is
not
paired at all, due to the failure of relation (\ref{p4}), as previously pointed
out.

\begin{cor} \label{coroll}
The coadjoint action of $U_{q}(h)$ on $\widetilde{BH_{q}(1)}$ is
\begin{eqnarray}
A \rhd \left( \begin{array}{ccc}
		1 & \alpha & \beta \\
		0 & \gamma & \delta \\
		0 & 0 & 1
	      \end{array} \right)   & = &
		     \left( \begin{array}{ccc}
				0 & 1 - \gamma & -\delta \\
				0 & 0 & 0 \\
				0 & 0 & 0
			    \end{array} \right)   \nonumber \\
 & & \nonumber \\
A^{\dagger } \rhd \left( \begin{array}{ccc}
		1 & \alpha & \beta \\
		0 & \gamma & \delta \\
		0 & 0 & 1
	      \end{array} \right)   & = &
		     \left( \begin{array}{ccc}
				0 & 0 & \alpha \\
				0 & 0 & \gamma - 1 \\
				0 & 0 & 0
			    \end{array} \right)   \nonumber \\
 & & \nonumber \\
N \rhd \left( \begin{array}{ccc}
		1 & \alpha & \beta \\
		0 & \gamma & \delta \\
		0 & 0 & 1
	      \end{array} \right)   & = &
		     \left( \begin{array}{ccc}
				0 & \alpha & 0 \\
				0 & 0 & -\delta \\
				0 & 0 & 0
			    \end{array} \right)   \nonumber \\
 & & \nonumber \\
H \rhd \left( \begin{array}{ccc}
		1 & \alpha & \beta \\
		0 & \gamma & \delta \\
		0 & 0 & 1
	      \end{array} \right)   & = & 0
\end{eqnarray}
\end{cor}
\proof
The above duality of $U_{q}(h)$ and $\widetilde{H_{q}(1)}$ allows the coadjoint
action of $U_{q}(h)$ on $\widetilde{H_{q}(1)}$ to be defined in terms of the
pairing
\eqn{coadactn}{X \rhd t^{i}\,_{j} = t^{k}\,_{l}\,
\langle X \,, (St^{i}\,_{k})\, t^{l}\,_{j} \rangle . }
According to theory described in \cite{Majex}, the braided group of function
algebra type is obtained from the quantum group of the same type, in the
quotient dual to the enveloping algebra, via a process of ``transmutation''. A
feature of this theory is
that the generators of the braided group obtained in this way, transform in the
same way as the generators of the quantum group, under the action of the dual
enveloping algebra. In the case of the Heisenberg group, this means that the
same set of coefficients $(\langle X \,, (St^{i}\,_{k})\,t^{l}\,_{j} \rangle )$
give the action of $U_{q}(h)$ on $\widetilde{BH_{q}(1)}$, as give its action on
$\widetilde{H_{q}(1)}$
\eqn{bractn}{X \rhd u^{i}\,_{j} = u^{k}\,_{l}\,
\langle X , (St^{i}\,_{k}) t^{l}\,_{j} \rangle . }
A straightforward application of this formula yields the actions quoted.
\endproof

Since a key motivation for introducing braided groups was their covariance
properties, it may be of interest at this stage to see explicitly why
$\widetilde{BH_{q}(1)}$ is $U_{q}(h)$-covariant, while $\widetilde{H_{q}(1)}$
is
not. Since the action chosen was the quantum coadjoint action, it will by
definition respect the coproduct structure of both groups. As previously
pointed
out, it is the product of $\widetilde{H_{q}(1)}$ which we expect not to be
covariant under this action. It is sufficient to demonstrate this with an
explicit example. Thus
\begin{eqnarray}
A^{\dagger } \rhd (\beta \delta - \delta \beta ) & = & A^{\dagger }
 \rhd (\beta \delta) - A^{\dagger } \rhd (\delta \beta ) \nonumber \\
   & = & (A^{\dagger }\,_{(1)} \rhd \beta ) (A^{\dagger }\,_{(2)} \rhd \delta )
-(A^{\dagger }\,_{(1)} \rhd \delta ) (A^{\dagger }\,_{(2)} \rhd \beta )
\nonumber \\
   & = & (e^{-\frac{\omega H}{4} } \rhd \beta ) (A^{\dagger } \rhd \delta )
+ (A^{\dagger } \rhd \beta ) (e^{\frac{\omega H}{4} } \rhd \delta )
\nonumber \\
   &   & \ \ \ \ \ \ \ \ -(e^{-\frac{\omega H}{4} } \rhd \delta )
(A^{\dagger } \rhd \beta ) - (A^{\dagger } \rhd \delta )
(e^{\frac{\omega H}{4} } \rhd \beta ) \nonumber \\
   & = & \beta (\gamma - 1) + \alpha \delta -\delta \alpha - (\gamma - 1) \beta
\nonumber \\
   & = & \alpha \delta -\delta \alpha \nonumber \\
   & = & \left\{ \begin{array}{ll}
			0 & \ \ \ \ \  \widetilde{H_{q}(1)} \\
			\omega (\gamma - 1) & \ \ \ \ \  \widetilde{BH_{q}(1)}
	 \end{array} \right.
\end{eqnarray}
Now,
\begin{eqnarray}
\beta \delta - \delta \beta & = & \left\{ \begin{array}{ll}
		-\frac{\omega }{2} \delta & \ \ \ \ \   \widetilde{H_{q}(1)} \\
		\omega \delta & \ \ \ \ \  \widetilde{BH_{q}(1)}
			\end{array} \right.
\end{eqnarray}
so that for $\widetilde{BH_{q}(1)}$
\begin{eqnarray}
A^{\dagger } \rhd (\omega \delta )
   & = & \omega (\gamma - 1) \nonumber \\
 & = & A^{\dagger } \rhd (\beta \delta - \delta \beta )
\end{eqnarray}
while for $\widetilde{H_{q}(1)}$
\begin{eqnarray}
A^{\dagger } \rhd (-\frac{\omega }{2} \delta ) & = &
    -\frac{\omega }{2} (\gamma - 1)  \nonumber \\
 & \not= & A^{\dagger } \rhd (\beta \delta - \delta \beta )
\end{eqnarray}
It can be seen that while the action of $A^{\dagger }$ preserves the commutator
in $\widetilde{BH_{q}(1)}$, this is not the case for $\widetilde{H_{q}(1)}$,
where the action of $A^{\dagger }$ on the two halves of the commutator give
different results: the product is not respected.

Proceeding in similar fashion, it can easily be shown that all commutators of
$\widetilde{BH_{q}(1)}$ are preserved by the action given in
Corollary~\ref{coroll}. By contrast, one other action on $\widetilde{H_{q}(1)}$
will be found to fail: that of $A$ on the commutator $[\alpha ,\beta ]$.

In fact formula (\ref{bractn}) works more generally: it is true for any braided
group
which maps to $\widetilde{BH_{q}(1)}$ (eg.\ $BM(3)$, $BT_{+}(h)$). According to
general theory \cite{Majrev}, any universal enveloping algebra can be defined
as
being generated by 1 and $2n^{2}$ indeterminates ($2\ n \times n$ matrices
$l^{+}$, $l^{-}$), modulo certain matrix relations. For standard R-matrices,
this is part of the FRT \cite{FRT} approach, but it is also valid in the
general
case. Following \cite{Majskyl} we first note that these generators can be
expressed in terms of the universal $R$-matrix
\eqn{ldef}{l^{+} = \CR^{(1)} \langle t \,, \CR^{(2)} \rangle ,\ \ \ \ \ \ \ \
l^{-} =  \langle t \,, S\CR^{(1)} \rangle \CR^{(2)}   }
where $\CR = \CR^{(1)} \otimes \CR^{(2)}$. The action of the universal
enveloping algebra on $B(R)$ is then given in \cite{Majex,Majskyl}
\eqn{lactn}{l^{+}\,^{k}\,_{l} \rhd u^{i}\,_{j} =R^{-1}\,^{i}\,_{m}\,^{k}\,_{a}
\,u^{m}\,_{n} \,R^{n}\,_{j}\,^{a}\,_{l}
\ \ \ \ \ \ \ \ l^{-}\,^{i}\,_{j} \rhd u^{k}\,_{l} =
R^{i}\,_{a}\,^{k}\,_{m} \,u^{m}\,_{n} \,R^{-1}\,^{a}\,_{j}\,^{n}\,_{l}   }
Careful comparison of the notation will show that this is, in fact, equivalent
to Equation~(\ref{bractn}), (using some of the properties of the pairing.)

For $U_{q}(h)$, $l^{+}$ and $l^{-}$ can be computed from~(\ref{ldef}), and are
found to be, explicitly
\eqn{explicl}{l^{+} = \left( \begin{array}{ccc}
				1 & 0  & -\frac{\omega }{2} N \\
				0 & e^{-\frac{\omega H}{2} } &
\omega e^{-\frac{\omega H}{4} } A \\
				0 & 0 & 1
			\end{array} \right) \ \ \ \ \ \ \ \  l^{-} =
\left( \begin{array}{ccc}
		1 & -\omega e^{-\frac{\omega H}{4} } A^{\dagger} &
\frac{\omega }{2} N \\
		0 & e^{\frac{\omega H}{2} } & 0 \\
		0 & 0 & 1
\end{array} \right)   }
Using Equation~(\ref{lactn}), we then compute the general action of $U_{q}(h)$
on $BM_{h}(3)$. In terms of the generators, it is
\begin{eqnarray}
A \rhd u & = & \left( \begin{array}{ccc}
			-u_{21} & u_{11} - u_{22} &
				-u_{23} - \frac{\omega }{2} u_{21} \\
			0 & u_{21} & 0 \\
			0 & u_{31} & 0
		\end{array} \right) \nonumber \\
\ \ \ & \ \ \ & \ \ \   \nonumber \\
A^{\dagger} \rhd u & = & \left( \begin{array}{ccc}
			0 & 0 & u_{12} + \frac{\omega }{2} u_{32} \\
			-u_{31} & -u_{32} & u_{22} - u_{33} \\
			0 & 0 & u_{32}
		\end{array} \right) \nonumber \\
\ \ \ & \ \ \ & \ \ \   \nonumber \\
N \rhd u & = & \left( \begin{array}{ccc}
			0 & u_{12} & 0 \\
			-u_{21} & 0 & -u_{23} \\
			0 & u_{32} & 0
		\end{array} \right) \nonumber \\
 & &   \nonumber \\
e^{\pm \frac{\omega H}{2} } \rhd u & = & \left( \begin{array}{ccc}
			u_{11} \mp \frac{\omega }{2} u_{31} & u_{12} \mp u_{32}
& u_{13} \mp \frac{\omega }{2}(u_{33} - u_{11}) - \frac{\omega^{2} }{4}
u_{31}\\
			u_{21} & u_{22} & u_{23} \pm \frac{\omega }{2} u_{21} \\
			u_{31} & u_{32} & u_{33} \pm \frac{\omega }{2} u_{31}
		\end{array} \right)
\end{eqnarray}
It is easy to see, by inspection, that these actions reduce to those already
given for $\widetilde{BH_{q}(1)}$, when the relevant quotient is taken.

All of the $U_{q}(h)$-modules which have thus far been considered have been
braided groups of function algebra type. It is also possible for $U_{q}(h)$ to
act on quantum or braided groups of enveloping algebra type. $U_{q}(h)$ does in
fact act upon itself, as an algebra only, via the quantum adjoint action. This
action
does not, however, respect its coalgebra structure. In analogy with the
function
algebra case, a braided version of $U_{q}(h)$ can be defined which has the same
algebra, but a modified coproduct, which will then be covariant under the
action \cite{MajalgQFT}. It may naturally be denoted $BU_{q}(h)$. See
\cite{MajalgQFT} for
details of the construction.
\begin{theorem} \label{isomorphism}
$\widetilde{BH_{q}(1)}$ is isomorphic to $BU_{q}(h)$.
\end{theorem}
\proof
Let the matrix $L$ be defined by
\eqn{Ldef}{L = l^{+} S l^{-} }
Then the identification $u = L$ always gives a homomorphism between the braided
groups of function and enveloping algebra types \cite{Majskyl}. In this case,
the map $u \longmapsto L$ will give a homomorphism $\widetilde{BH_{q}(1)}
\longmapsto BU_{q}(h)$. For the Heisenberg group, $L$ is, explicitly
\eqn{Lexplic}{L = \left( \begin{array}{ccc}
	    1 & \omega e^{ -\frac{\omega H}{4} } A^{\dagger } & -\omega N   \\
	    0 & e^{ -\omega H} & \omega e^{ -\frac{\omega H}{4} } A   \\
	    0 & 0 & 1
		\end{array} \right) }
It can easily be verified directly that this does indeed give an algebra
homomorphism, (remembering that by definition $BU_{q}(h)$ has the same algebra
(on the generators) as $U_{q}(h)$.) Also, the map $u \longmapsto L$ is
obviously
one-to-one when $u$ is taken to be the matrix of generators in the quotient
$\widetilde{BH_{q}(1)}$. Thus the identifications
\eqn{ident}{\alpha = \omega e^{ -\frac{\omega H}{4} } A^{\dagger },
\ \ \ \ \ \ \ \ \beta = -\omega N,\ \ \ \ \ \ \ \
\gamma = e^{ -\omega H},\ \ \ \ \ \ \ \
\delta = \omega e^{ -\frac{\omega H}{4} } A  }
define an isomorphism between $\widetilde{BH_{q}(1)}$ and $BU_{q}(h)$.
\endproof

This isomorphism can be exploited to derive the coalgebra structure (etc.)\ of
$BU_{q}(h)$ directly from that of $\widetilde{BH_{q}(1)}$. The coalgebra and
antipode are
\begin{eqnarray}
\underline{\Delta }A & = & A \otimes e^{ \frac{\omega H}{4} }
+ e^{ -3 \frac{\omega H}{4} } \otimes A   \nonumber \\
\underline{\Delta }A^{\dagger } & = & A^{\dagger } \otimes e^{ -3 \frac{\omega
H}{4} }
+ e^{ \frac{\omega H}{4} } \otimes A^{\dagger }   \nonumber \\
\underline{\Delta }N & = & N \otimes 1 + 1 \otimes N
- \omega e^{-\frac{\omega H}{4}} A^{\dagger } \otimes e^{-\frac{\omega H}{4}} A
\nonumber \\
\underline{\Delta }H & = & H \otimes 1 + 1 \otimes H
\end{eqnarray}
\eqn{BUqhantipode}{\uS(A) = -e^{ \frac{\omega H}{2} } A,\ \ \ \ \
\uS(A^{\dagger }) = -e^{ \frac{\omega H}{2} } A^{\dagger },\ \ \ \ \
\uS(N) =N - e^{ \frac{\omega H}{2} } A^{\dagger } A,\ \ \ \ \
\uS(H) = -H .}
The coproduct on $BU_{q}(h)$ is seen to be considerably less symmetric than
that
on $U_{q}(h)$. A more complicated coalgebra structure is often observed in
braided universal enveloping algebras, as compared to their quantum
counterparts. This can be partially accounted for by the fact that the
universal $R$-matrix
is just 1 for the braided group, and some of the information which used to be
contained (in the quantum group) in this quasi-triangular structure, is now
``transferred'' to the coalgebra structure. In $BU_{q}(h)$, for example, a
similarity can be observed between the extra term in the coproduct of $N$ and
part of the $R$-matrix of $U_{q}(h)$.

The action of $U_{q}(h)$ on $BU_{q}(H)$ can also be derived directly from the
isomorphism. It is in fact the same as the adjoint action of $U_{q}(h)$ on
itself, as defined by the formula \cite[Section 6]{Majrev}
\eqn{adjactn}{h \rhd a = h_{(1)} a Sh_{(2)} }
where $\Delta h = h_{(1)} \otimes h_{(2)}$. Explicitly
\begin{eqnarray}
\label{action}
A \rhd \left( \begin{array}{cc}
		A^{\dagger } & N \\
		H & A
	\end{array} \right)  & = & \left( \begin{array}{cc}
		e^{-\frac{\omega H}{4} } [H] & e^{ -\frac{\omega H}{4} } A  \\
		0 & 0
				   \end{array} \right)   \nonumber  \\
 &  &    \nonumber  \\
A^{\dagger } \rhd \left( \begin{array}{cc}
		A^{\dagger } & N \\
		H & A
	\end{array} \right)  & = & \left( \begin{array}{cc}
		0 & -e^{ -\frac{\omega H}{4} } A^{\dagger }  \\
		0 & -e^{ -\frac{\omega H}{4} } [H]
				   \end{array} \right)   \nonumber  \\
 &  &    \nonumber  \\
N \rhd \left( \begin{array}{cc}
		A^{\dagger } & N \\
		H & A
	\end{array} \right)  & = & \left( \begin{array}{cc}
		A^{\dagger } & 0 \\
		0 & -A
				   \end{array} \right)   \nonumber  \\
 &  &    \nonumber  \\
H \rhd \left( \begin{array}{cc}
		A^{\dagger } & N \\
		H & A
	\end{array} \right)  & = & 0
\end{eqnarray}
The quantum adjoint action is always respected by the product. It is also
respected by the modified coproduct of $BU_{q}(h)$ (since it is isomorphic to
the coadjoint action on the function algebra.) Thus $BU_{q}(h)$ is
$U_{q}(h)$-covariant.

To complete the description of $BU_{q}(h)$, it only remains to give the
braiding. This can be computed using the isomorphism in Theorem 4.3.
\begin{eqnarray}
\label{univpsi}
\Psi (A^{\dagger } \otimes A^{\dagger }) & = & A^{\dagger } \otimes A^{\dagger
} \nonumber  \\
\Psi (A^{\dagger } \otimes N) & = & N \otimes A^{\dagger }
- \omega e^{ -\frac{\omega H}{4} } A^{\dagger } \otimes e^{ -\frac{\omega H}{4}
} [H] \nonumber  \\
\Psi (A^{\dagger } \otimes A) & = & A \otimes A^{\dagger }
- \omega e^{ -\frac{\omega H}{4} } [H] \otimes e^{ -\frac{\omega H}{4} } [H]
\nonumber  \\
\Psi (N \otimes A^{\dagger }) & = & A^{\dagger } \otimes N \nonumber  \\
\Psi (N \otimes N) & = & N \otimes N - \omega e^{ -\frac{\omega H}{4} }
A^{\dagger } \otimes e^{ -\frac{\omega H}{4} } A  \nonumber  \\
\Psi (N \otimes A) & = & A \otimes N
- \omega e^{ -\frac{\omega H}{4} } [H] \otimes e^{ -\frac{\omega H}{4} } A
\nonumber  \\
\Psi (A \otimes A^{\dagger }) & = & A^{\dagger } \otimes A \nonumber  \\
\Psi (A \otimes N) & = & N \otimes A \nonumber  \\
\Psi (A \otimes A) & = &  A \otimes A
\end{eqnarray}
where $[H] = \frac{1}{\omega }(e^{\frac{\omega H}{2}}-e^{-\frac{\omega
H}{2}})$.

\section{The n-body System of Braided Harmonic Oscillators}
\label{sec-app}

In the Introduction to this paper, it was mentioned that $U_{q}(h)$ was
essentially the `quantum group version' of the ordinary quantum harmonic
oscillator. $BU_{q}(h)$, then, can similarly be considered as describing a
braided harmonic oscillator. These oscillators both have the same algebra: the
difference between them is that the latter is to be viewed as existing within a
braided category, in the sense that was explained in the last section. This
difference only becomes apparent when we consider a system of more than one
oscillator. A system of n braided oscillators will be represented on the n-fold
(braided) tensor product algebra \cite{MajalgQFT,Majex}. This algebra is as
defined in Equation~(\ref{btp}), iterated n times. We denote it
$BU_{q}(h)^{\underline{\otimes } n}$. (It will be convenient, throughout this
section, to distinguish between braided and unbraided tensor products by using
an underlined symbol for the braided product.) In this Section,
$BU_{q}(h)^{\underline{\otimes } n}$ and its representations will be studied
with the above application in mind. We will attempt to define braided analogues
of all the usual quantum-mechanical concepts: Fock space, an inner product, and
time-evolution.

We begin by recalling that an algebra isomorphism exists between $U_{q}(h)$ and
$BU_{q}(h)$, so that for a single oscillator, the braided and unbraided sytems
are equivalent. However, the braiding would {\em a-priori} appear to cause
significant
differences in the tensor product algebras $U_{q}(h)^{\otimes n}$ and
$BU_{q}(h)^{\underline{\otimes } n}$. The braided and unbraided n-body systems
would therefore appear quite different, despite the identical nature of the
individual oscillators which comprise them. The algebra of
$U_{q}(h)^{\otimes n}$ is just the usual n-body harmonic oscillator algebra,
with quantum group modification,
\eqn{nbodalg1}{[A_{i}, A_{j}] = 0 \ \ \ \ \
[A^{\dagger }_{i}, A^{\dagger }_{j}] = 0\ \ \ \ \
[A_{i}, A^{\dagger }_{j}] =  \delta_{ij} [H_{i}] }
\eqn{nbodalg2}{[N_{i}, A_{j}] = -\delta_{ij} A_{i} \ \ \ \ \
[N_{i}, A^{\dagger }_{j}] = \delta_{ij} A^{\dagger } \ \ \ \ \
[N_{i}, N_{j}] = 0 \ \ \ \ \ H_{i}\ {\rm central}  }
By contrast, the algebra of $BU_{q}(h)^{\underline{\otimes } n}$ is much more
complicated, due to the braiding. It is calculated by repeated use of
Equation~(\ref{btp}), with $\Psi $ as given in (\ref{univpsi}), to give
\begin{eqnarray}
\label{nbalg}
\,[\Abs{i}, \Abs{i}] & = & 0  \nonumber  \\
\,[\Adbs{i}, \Adbs{j}] & = & 0
\nonumber  \\
\,[\Abs{i}, \Adbs{j}] & = & \left\{ \begin{array}{ll}
	\,\omega \,\eHn{i}{4}\, [\Hbs{i}]\, \eHn{j}{4}\, [\Hbs{j}]  & i<j  \\
	\,[\Hbs{i}]  &  i=j  \\
	\,0 &  i>j
				\end{array} \right.
\nonumber  \\
\,[\Nbs{i}, \Abs{j}] & = &  \left\{ \begin{array}{ll}
	\,0  &  i<j  \\
	\,-\Abs{i}  & i=j  \\
	\,\omega \, \eHn{j}{4}\, [\Hbs{j}]\, \eHn{i}{4}\, \Abs{i}  &  i>j
				\end{array} \right.
\nonumber  \\
\,[\Nbs{i}, \Adbs{j}] & = &  \left\{ \begin{array}{ll}
	\,\omega \, \eHn{i}{4}\, \Adbs{i}\, \eHn{j}{4}\, [\Hbs{j}]  &  i<j  \\
	\,\Adbs{i}  &  i=j  \\
	\,0  & i>j
				\end{array} \right.
\nonumber  \\
\,[\Nbs{i}, \Nbs{j}] & = &  \left\{ \begin{array}{ll}
	\,\omega \,\eHn{i}{4}\, \Adbs{i}\, \eHn{j}{4}\, \Abs{j}  &  i<j  \\
	\,0  & i=j
				\end{array} \right.
\nonumber  \\
\,[\Hbs{i}, \bar{Y}_{j}] & = & 0 \ \ \ \ \ \forall \:\bar{Y} = \Ab, \Adb, \Nb,
\Hb
\end{eqnarray}
where $A_{i}$\ (\Abs{i}) is that element of \nuqh\ (\nb ) which has $A$ in the
$i^{{\rm th}}$ place and $1$ elsewhere; and equivalently for the other
generators. This definition will hold throughout this section, except where
otherwise explicitly noted. Elements of \nb are barred to make explicit the
distinction between braided and unbraided tensor products. The notation
\eqn{notdef}{[H] =
\frac{1}{\omega }\,(e^{\frac{\omega H}{2}} - e^{-\frac{\omega H}{2}})  }
will also be adopted universally.

Note that the {\em order of labelling\/} matters in \nb. This, though a natural
consequence of the braiding, would seem a little odd physically: exchanging the
labels of two oscillators changes the {\em value\/} (and not just the sign) of
some of their commutators. However, this peculiarity of braided statistics may
not actually be manifest in physical states of the system, due to the existence
of an isomorphism, (which will be demonstrated later this section),
between braided and unbraided tensor products.

Having written down the n-fold braided tensor product algebra, the next task is
to find a suitable Fock Space representation for it. Such representations of
$U_{q}(h)$ are already known: a generic definition is given in \cite{GS}
\begin{eqnarray}
\label{repdef}
A\,|r \rangle & = & \half{\hbar } \sqrt{r}\,|r-1 \rangle  \nonumber \\
\Ad\,|r \rangle & = & \half{\hbar } \sqrt{r + 1}\,|r+1 \rangle  \nonumber \\
H\,|r \rangle & = & \hbar\,|r \rangle  \nonumber \\
N\,|r \rangle & = & (n^{\prime } + r)\,|r \rangle
\end{eqnarray}
where $\{|r \rangle \}^{\infty }_{r=0}$ is an orthonormal basis; and the irrep
is labelled by the eigenvalues of the central elements $H$ and
$C(=[H]N - \Ad A)$
(being $\hbar $ and $[\hbar ] n^{\prime }$ respectively). It is obvious from
the
above definition that any state $|r \rangle $ can be expressed in terms of the
vacuum as follows
\eqn{fock}{|r \rangle = \frac{(A^{\dagger })^{r}}{
[\hbar ]^{\frac{r}{2}}\,\sqrt{r!} }\:|0 \rangle }
and that therefore the given irreps are indeed Fock Space representations.
It is quite easy to generalise the representation from $1$ to n oscillators for
$U_{q}(h)$. A tensor product of the single oscillator bases will form a Fock
Space representation of \nuqh.
\begin{eqnarray}
\label{unbrfockrep}
|r_{1}, r_{2}, \ldots ,r_{n} \rangle & = &
|r_{1} \rangle\,\otimes\,|r_{2}\,\rangle\,\otimes \cdots \otimes\,|r_{n}
\rangle
\nonumber \\
  & = & \frac{(\Ads{1})^{r_{1}} (\Ads{2})^{r_{2}} \cdots (\Ads{n})^{r_{n}} }
{[\hbar_{1}]^{\frac{r_{1}}{2}} \cdots [\hbar_{n}]^{\frac{r_{n}}{2}}
\,\sqrt{r_{1}! r_{2}! \cdots r_{n}!} }
\:|0 \rangle \otimes |0 \rangle \otimes \cdots \otimes |0 \rangle
\end{eqnarray}
Note that the individual oscillators need not all be represented by the same
irrep. To maintain generality, the eigenvalues of $H_{i}$ and $C_{i}$ in the
$i^{th}$ oscillator are denoted by $\hbar_{i}$ and $[\hbar_{i}] n^{\prime}_{i}
$
respectively. ``Ordinary" quantum mechanics can be obtained from this more
general `quantum group version', by setting $\hbar_{i} = \hbar$ and
$n^{\prime}_{i} = 0$, $\forall i$. The system therefore consists of n
independent harmonic oscillators, each of which has fixed, though possibly
different, values of Planck's constant and the vacuum expectation value of the
number operator.

For the braided algebra, the situation is not so trivial. At the single
oscillator level, $BU_{q}(h)$ obviously has the same representations as
$U_{q}(h)$, since it has the same algebra. An n-fold basis can be constructed
in an
analogous way to that for $U_{q}(h)^{\otimes n}$, by taking now the
{\em braided\/} tensor product of the individual bases. Thus
\eqn{brtens}{|r_{1}, \ldots ,r_{n} \rangle =
|r_{1} \rangle\,\utens\,|r_{2} \rangle\,\utens \cdots \utens\,|r_{n} \rangle }
However, this cannot be written as a Fock Space in any obvious manner, since
because of the braiding the \Ad's will not necessarily commute with
the states. To find the action of any of the operators on the states as defined
above, we must first compute the braiding between the generators and the
(single-oscillator) states.

The braiding can be calculated from the universal $R$-matrix, given in
Equation~(\ref{algformRmat}), and the known actions of $U_{q}(h)$ on both
$BU_{q}(h)$ and the representation (Equations~(\ref{action}) and (\ref{repdef})
respectively), using the standard formula \cite[section 7]{Majrev}:
\eqn{algpsi}{\Psi (v\,\utens\,w) = \sum \CR^{(2)}\!\rhd\!w\
\utens\:\CR^{(1)}\!\rhd\!v \ \ \ \ \ \ (\CR = \sum \CR^{(1)} \otimes
\CR^{(2)} )   }
The results are as follows
\begin{eqnarray}
\Psi (A\,\utens\,|r_{i} \rangle ) & = & \eh{i}{2}\,|r_{i} \rangle\,\utens\,A
\nonumber \\
\Psi (\Ad\,\utens\,|r_{i} \rangle ) & = & \omega\,\ehn{i}{2}\,\half{\hbar_{i}}
\,\sqrt{r_{i} +1}\,|r_{i} +1 \rangle\,\utens\,e^{-\frac{\omega }{4}H}[H]
\;+\;\ehn{i}{2} |r_{i} \rangle\,\utens\,\Ad \nonumber  \\
\Psi (N\,\utens\,|r_{i} \rangle ) & = & \omega\,\eh{i}{4}\,\half{\hbar_{i}}
\sqrt{r_{i} +1}\,|r_{i} +1 \rangle\,\utens\,e^{-\frac{\omega }{4}H} A
\;+\;|r_{i} \rangle\,\utens\,N  \nonumber  \\
\Psi (H\,\utens\,|r_{i} \rangle ) & = & |r_{i} \rangle\,\utens\,H
\end{eqnarray}

We are now in a position to calculate the action of arbitrary tensor products
of
operators on the braided space. To calculate such an action, each operator must
be braided through all the states until it reaches the one on which it should
act; only when this has been done for all operators should the actions be
computed. Proceeding in this fashion, we find the action of $A^{\dagger }_{i}$
on a general n-fold state to be
\begin{eqnarray}
\Adbs{i}\,|r_{1}, \ldots , r_{n} \rangle & = & \omega\,\ehn{i}{4}\,[\hbar_{i}]
\sum_{j=1}^{i-1} \{\esumhn{1}{j-1}{2}\,\ehn{j}{4}\,\half{\hbar_{j}}
\sqrt{r_{j} +1}\,|r_{1}, \ldots , r_{j} +1, \ldots , r_{n} \rangle \}
\nonumber  \\
  &  & \ \ \ \ \ \ \ \ \ \  + \esumhn{1}{i-1}{2}\,\half{\hbar_{i}}
\sqrt{r_{i} +1}\,|r_{1}, \ldots , r_{i} +1, \ldots , r_{n} \rangle
\end{eqnarray}
If we postulate the existence of some operators \Xds{i}\ which have the
following action on a general state
\eqn{Xdaction}{\Xds{i}\,|r_{1}, \ldots , r_{n} \rangle = \half{\hbar_{i}}
\sqrt{r_{i} +1}\,|r_{1}, \ldots , r_{i} +1, \ldots , r_{n} \rangle }
then it is clear that \Adbs{i}\  can be written as a sum of such operators
\eqn{AdsumXd}{\Ads{i} = \esumHn{1}{i-1}{2}\,\Xds{i} + \omega\,\eHn{i}{4}\,
[\bar{H}_{i}] \sum_{j=1}^{i-1} \{\esumHn{1}{j-1}{2}\,\eHn{j}{4}\,\Xds{j} \} }
This relation can then be inverted to give a definition of \Xds{i}\ in terms of
the \Adbs{i}'s
\eqn{XdsumAd}{\Xds{i} = \esumH{1}{i-1}{2}\,\Adbs{i} - \omega\,[\bar{H}_{i}]\,
\eHn{i}{4}
\sum_{j=1}^{i-1} \{ \esumH{1}{j-1}{2}\,\eH{j}{4}\,\esumHn{j+1}{i-1}{2}\,
\Adbs{j} \}}
It is clear from this last equation that the \Xd's are properly defined
operators in \nb, although the subscript notation must be understood
differently
from the usual convention. Written out as a tensor
product, \Xds{i}\ does {\em not\/} consist of \Xd\ in the $i^{th}$ place and 1
elsewhere; it is a sum of such elements, and will contain elements different
from 1 in all places up to and including the $i^{th}$.

In effect, \Xds{i}\ can be considered as a `diagonalised version' of \Adbs{i} .
It is clear from the action of the \Xd 's on the representation that it is
they,
{\em and not the \Ad 's \/}, which are the physical creation operators on the
braided space. The existence of this diagonalisation implies that the braided
representation of \nb\ (Equation~(\ref{brtens})\,) is a Fock Space; explicitly
\eqn{brfock}{|r_{1}, \ldots ,r_{n} \rangle = \frac{(\Xds{1})^{r_{1}} \cdots
(\Xds{n})^{r_{n}}}{[\hbar_{1}]^{\frac{r_{1}}{2}} \cdots
[\hbar_{n}]^{\frac{r_{n}}{2}} \sqrt{r_{1}! \cdots r_{n}!} }\:|0 \rangle\,\utens
\cdots \utens\,|0
\rangle  }
The similarity of this formula to Equation~(\ref{unbrfockrep}) suggests the
possible existence of an isomorphism between the two Fock Spaces.

\begin{theorem}
\label{th-tensiso}
The algebras \nb and \nuqh are isomorphic.
\end{theorem}
\proof
We begin by `diagonalising' \Ab, as it is the diagonalised versions of \Adb\
and
\Ab\ which we suspect (due to the similarity of formulas (\ref{unbrfockrep})
and
(\ref{brfock}) ) may be isomorphic to the unbraided tensor algebra. The
action of \Ab\ on a general state is
\eqn{Aactn}{\Abs{i}\,|r_{1}, \ldots ,r_{n} \rangle = \esumh{1}{i-1}{2}\,
\half{\hbar_{i}}\,\sqrt{r_{i}}\,|r_{1}, \ldots, r_{i}-1, \ldots r_{n} \rangle }
{}From this action it is clear that if \Xs{i}\ is defined as follows in terms
of
\Abs{i}
\eqn{XitoA}{\Xs{i} = \esumHn{1}{i-1}{2}\,\Abs{i} }
then it will have the desired diagonal action on a general state
\eqn{Xdiagactn}{\Xs{i}\,|r_{1}, \ldots, r_{n}\rangle = \half{\hbar_{i}}\,
\sqrt{r_{i}}\,|r_{1}, \ldots, r_{i}-1, \ldots, r_{n}\rangle }
We now proceed to compute the commutation relations of \Xds{i}\ and \Xs{i} , as
determined by the braided relations between \Adbs{i}\ and \Abs{i}, given in
Equation~(\ref{nbalg}). Since \Xds{i}\ is a function only of \Adbs{i}, we have
immediately
\eqn{Xdcomm}{[\Xds{i},\Xds{j}] = 0 }
and similarly, noting that \Xs{i}\ likewise is a function only of \Abs{i}
\eqn{Xcomm}{[\Xs{i},\Xs{j}] = 0 }
Since the commutator $[\Abs{i},\Adbs{j}]$ has different values depending on the
order of i and j, it will be convenient to calculate the three cases for
$[\Xs{i},\Xds{j}]$ separately.
\begin{eqnarray}
\,[\Xs{i}, \Xds{j}]_{i>j} & = &
[\esumHn{1}{i-1}{2}\,\Abs{i}\:,\:\esumH{1}{j-1}{2}
\,\Adbs{j}   \nonumber  \\
 & & \ \ \ \ \ \ - \omega\,[\Hbs{j}]\,\eHn{j}{4} \sum_{k=1}^{j-1}
\{\esumH{1}{k-1}{2}\,\eH{k}{4}\,\esumHn{k+1}{j-1}{2}\,\Adbs{j}\} ]  \nonumber
\\
 & = & 0  \\
 & & \nonumber \\
\,[\Xs{i}, \Xds{i}] & = & [\esumHn{1}{i-1}{2}\,\Abs{i}\:,\;\esumH{1}{i-1}{2}
\,\Adbs{i}   \nonumber  \\
 & & \ \ \ \ \ \ - \omega\,[\Hbs{i}]\,\eHn{i}{4} \sum_{j=1}^{i-1}
\{\esumH{1}{j-1}{2}\,\eH{j}{4}\,\esumHn{j+1}{i-1}{2}\,\Adbs{k}\} ]  \nonumber
\\
 & = & [\Abs{i},\Adbs{i}]   \nonumber \\
 & = & [\Hbs{i}]   \\
 & & \nonumber \\
\label{ilessj}
\,[\Xs{i}, \Xds{j}]_{i<j} & = &
[\esumHn{1}{i-1}{2}\,\Abs{i}\:,\:\esumH{1}{j-1}{2}\,\Adbs{j}   \nonumber  \\
 & & \ \ \ \ \ \ - \omega\,[\Hbs{j}]\,\eHn{j}{4} \sum_{k=1}^{j-1}
\{\esumH{1}{k-1}{2}\,\eH{k}{4}\,\esumHn{k+1}{j-1}{2}\,\Adbs{k}\} ]  \nonumber
\\
 & = & \esumHn{1}{i-1}{2}\,\esumH{1}{i-1}{2}\:[\Abs{i}\:,\:\esumH{i}{j-1}{2}\,
\Adbs{j}  \nonumber \\
 & & \ \ \ \ \ \ - \omega\,[\Hbs{j}]\,\eHn{j}{4} \sum_{k=i}^{j-1}
\{\esumH{i}{k-1}{2}\,\eH{k}{4}\,\esumHn{k+1}{j-1}{2}\,\Adbs{k}\} ]  \nonumber
\\
 & = & \esumH{i}{j-1}{2}\,[\Abs{i},\Adbs{j}] - \omega\,[\Hbs{j}]\,\eHn{j}{4}\,
\eH{i}{4}\,\esumHn{i+1}{j-1}{2}\,[\Abs{i},\Adbs{i}]  \nonumber  \\
 & & \ \ \ \ \ \ - \omega\,[\Hbs{j}]\,\eHn{j}{4} \sum_{k=i+1}^{j-1}
\{\esumH{i}{k-1}{2}\,\eH{k}{4}\,\esumHn{k+1}{j-1}{2}\,[\Abs{i},\Adbs{k}] \}
\nonumber \\
 & = & \esumH{i}{j-1}{2}\,\omega\,\eHn{i}{4}\,[\Hbs{i}]\,\eHn{j}{4}\,[\Hbs{j}]
\nonumber  \\
 & & \ \ \ \ \ \ -
\omega\,[\Hbs{j}]\,\eHn{j}{4}\,\eH{i}{4}\,\esumHn{i+1}{j-1}{2}\,[\Hbs{i}]
\nonumber  \\
 & & \ \ \ \ \ \  - \omega\,[\Hbs{j}]\,\eHn{j}{4} \sum_{k=i+1}^{j-1}
\{\esumH{i}{k-1}{2}\,  \nonumber  \\
 & & \ \ \ \ \ \ \eH{k}{4}\,\esumHn{k+1}{j-1}{2}\,\omega\,\eHn{i}{4}\,
[\Hbs{i}]\,\eHn{k}{4}\,[\Hbs{k}]\}  \nonumber  \\
 & = & \omega\,\ersumHn{i}{j}{4}\,[\Hbs{i}][\Hbs{j}]\{\esumH{i}{j-1}{2} -
\eH{i}{2}
\,\esumHn{i+1}{j-1}{2}    \nonumber  \\
 & & \ \ \ \ \ \  - \sum_{k=i+1}^{j-1}
\{\esumH{i}{k-1}{2}\,\esumHn{k+1}{j-1}{2}
\,(\eH{k}{2} -\eHn{k}{2}) \} \}   \nonumber  \\
 & = & \omega\,\ersumHn{i}{j}{4}\,[\Hbs{i}][\Hbs{j}]\{\esumH{i}{j-1}{2} -
\eH{i}{2}
\,\esumHn{i+1}{j-1}{2}    \nonumber  \\
 & & \ \ \ \ \ \  - \sum_{k=i+1}^{j-1} \{\esumH{i}{k}{2}\,\esumHn{k+1}{j-1}{2}
\}  \nonumber  \\
 & & \ \ \ \ \ \ \ \  + \sum_{k=i+1}^{j-1} \{\esumH{i}{k-1}{2}\,
\esumHn{k}{j-1}{2} \} \} \nonumber  \\
 & = & \omega\,\ersumHn{i}{j}{4}\,[\Hbs{i}][\Hbs{j}]\{\esumH{i}{j-1}{2} -
\eH{i}{2}
\,\esumHn{i+1}{j-1}{2}    \nonumber  \\
 & & \ \ \ \ \ \  - \esumH{i}{j-1}{2} + \eH{i}{2}\,\esumHn{i+1}{j-1}{2} \}
v\nonumber  \\
 & = & 0
\end{eqnarray}
In Equation~(\ref{ilessj}), the second-last equality is obtained by writing out
both sums; all but one term in each cancel pairwise with terms in the other.
So,
combining the above three equations, we have
\eqn{finalXcomm}{[\Xs{i}, \Xds{j}] = \delta_{ij} [\Hbs{i}] }
which is the same as the {\em unbraided} commutator between $A_{i}$ and
\Ads{j},
asuming that \Hbs{i} is identified with $H_{i}$. This is certainly possible, as
the central elements of two algebras can always be identified.

By the same argument, the quadratic casimirs $C_{i}$ and $\bar{C}_{i}$ can also
be identified, and from this the element (\Mbs{i}) of \nb\ which is isomorphic
to $N_{i}$ can be found. It is
\eqn{Mdef}{\Mbs{i} = \Nbs{i} + \frac{1}{[\Hbs{i}]} \{ \Xds{i} \Xs{i} - \Adbs{i}
\Abs{i} \} \ \ \ \ \ \ (= \frac{1}{[\Hbs{i}]} \{ \bar{C}_{i} + \Xds{i} \Xs{i}
\}
) }
It is easily seen (by inspection) that \Mbs{i} will act on the representation
as a `diagonalised version' of \Nbs{i}
\eqn{Mact}{\Mbs{i} |r_{1}, \ldots ,r_{n} \rangle = (n^{\prime}_{i} + r_{i} )
|r_{1}, \ldots ,r_{n} \rangle }

Thus the identifications
\begin{eqnarray}
\Xds{i} & = & \Ads{i}  \nonumber \\
\Xs{i} & = & A_{i}  \nonumber \\
\Hbs{i} & = & H_{i}  \nonumber \\
\Mbs{i} & = & N_{i}
\end{eqnarray}
define an isomorphism between \nb and \nuqh.
\endproof

The form of the isomorphism makes it clear that the braided and unbraided Fock
Spaces are indeed equivalent. The strange dependence of the braided generators
on the order of labeling {\em does not appear\/} in the physical creation and
annihilation operators, \Xs{i}\ and \Xds{i}. Physical states of the system will
therefore be unaffected by the braided statistics.

Having found a Fock Space representation for \nb, we now turn to the question
of
whether a $*$-structure can be found which will allow an inner product to be
defined between the Fock Space states.

\begin{cor}
A $*$-structure exists on \nb, which is consistent with orthonormality of the
Fock Space basis.
\end{cor}
\proof
We begin by considering the 1-dimensional problem.
We want our representation to form an orthonormal basis for the Fock Space, so
we define the inner product
\eqn{ipdefinition}{(|n \rangle , |m \rangle ) = \delta_{n,m}  }
This then defines the $*$-structure of the generators, through the property
\eqn{starprop}{(|n \rangle,a|m \rangle ) = ( a^{*}|n \rangle,
|m \rangle )   }
It can immediately be seen that the $*$-structure
\eqn{starstruc}{(\Ad )^{*} = A,\ \ \ \ \ \ A^{*} = \Ad,\ \ \ \ \ \
N^{*} = N,\ \ \ \ \ \ H^{*} = H  }
is consistent with the inner product as defined. This is a known $*$-structure
of $U_{q}(h)$: it can easily be checked that $*$ as defined in
Equation~(\ref{starstruc}) extends as an anti-algebra map, if $\omega $ is
assumed to be real.

This $*$-structure must now be extended to encompass the n-dimensional system.
This can easily be done for \nuqh. Clearly the inner product
\eqn{ndimipdef}{(|r_{1}, \ldots ,r_{n} \rangle, |s_{1}, \ldots ,s_{n} \rangle )
= \delta_{r_{1}, s_{1}} \delta_{r_{2}, s_{2}} \cdots \delta_{r_{n}, s_{n}}  }
immediately implies
\eqn{ndimstarstruc}{A_{i}^{*} = \Ads{i},\ \ \ \ \ \ (\Ads{i})^{*} = A_{i},
\ \ \ \ \ \ N_{i}^{*} = N,\ \ \ \ \ \ H_{i}^{*} = H_{i}   }
by exactly the same argument as was used for the 1-dimensional problem. It is
also immediately clear that $*$ extends as an anti-algebra map for these
operators, and therefore for all elements of \nuqh.

The isomorphism can now be invoked to transfer this result directly to \nb. The
inner product as defined in Equation~(\ref{ndimipdef}) will be consistent with
the following $*$-structure on the `diagonalised' generators
\eqn{diagstar}{\Xs{i}^{*} = \Xds{i},\ \ \ \ \ \ (\Xds{i})^{*} = \Xs{i},\ \ \ \
\ \ \Mbs{i}^{*} =\Mbs{i},\ \ \ \ \ \ \Hbs{i}^{*} = \Hbs{i} }

This then implies a rather complicated $*$-structure for the usual generators
\begin{eqnarray}
\Abs{i}^{*} & = & \esumHnd{1}{i-1} \Adbs{i} - \omega\,[\Hbs{i}]\,\eHn{i}{4}
\sum_{j=1}^{i-1} \{\esumHnd{1}{j-1}\,e^{\frac{3 \omega }{4} \Hbs{j}} \Adbs{j}
\}
\nonumber  \\
(\Adbs{i})^{*} & = & \esumHnnd{1}{i-1} \Abs{i} + \omega\,[\Hbs{i}]\,\eHn{i}{4}
\sum_{j=1}^{i-1} \{\esumHnnd{1}{j-1}\,\eHn{j}{4} \Abs{j} \}  \nonumber  \\
\Nbs{i}^{*} & = & \Nbs{i} -\omega \sum_{j=1}^{i-1}
(\esumHnnd{j+1}{i-1} \ersumHn{i}{j}{4}\,\Adbs{i} \Abs{i})  \nonumber  \\
  &  & + \,\omega \sum_{j=1}^{i-1} ( \esumHnd{j+1}{j-1}\,\eHn{i}{4}\,
 e^{\frac{3 \omega }{4} \Hbs{j}} \Adbs{i} \Abs{j} ) \nonumber  \\
  & & \!\!\!\!\!\!\!\! -\, \omega^{2} \,[\Hbs{i}]\,\eHn{i}{2}
(\sum_{j=1}^{i-1} \esumHnd{1}{j-1}
e^{\frac{3 \omega }{4} \Hbs{j}} \Adbs{j})(\sum_{k=1}^{i-1} \esumHnnd{1}{k-1}\,
\eHn{k}{4} \Abs{k} )
\end{eqnarray}
\endproof

The last concept needed in order to fully define the quantum mechanics of
braided harmonic oscillators is their time evolution. It would be nice to be
able to describe this time evolution in terms of an action of $U_{q}(h)$ on
$BU_{q}(h)$, since this would ensure that all braided group maps were
preserved.
We therefore search for an element of $U_{q}(h)$ whose action on $BU_{q}(h)$
produces a physically reasonable time evolution.

An obvious choice is the quadratic casimir: $C=[H]N -\Ad A$, since being
central
it will preserve the symmetry of the system during evolution. Its actions on
the
generators of $BU_{q}(h)$ are as follows
\eqn{casact}{C \rhd \left( \begin{array}{cc}
				\Ad & N \\
				 H  & A
			\end{array} \right) = \left( \begin{array}{cc}
					0 & e^{-\frac{\omega }{2}H} [H]  \\
					0 & 0
					\end{array} \right)   }
This does not provide a very physical time evolution, since if $A$ and \Ad\ do
not evolve in time, neither will the states of the system. Only $N$ evolves,
and
this is an operator which should {\em not} evolve in time, as this would imply
change in the irrep describing the system.

Discarding the casimir as an option, then, another possibility is to choose a
primitive element of $U_{q}(h)$. This would have the advantage of giving
equations of motion very similar to the Heisenberg picture. Since $H$ has a
zero
action on all elements, we shall consider the action of $N$. It is, as was
given
in Equation~(\ref{action})
\eqn{Nact}{N \rhd \left( \begin{array}{cc}
				\Ad & N \\
				 H  & A
			\end{array} \right) = \left( \begin{array}{cc}
					\Ad & 0  \\
					 0 & -A
					\end{array} \right)   }
This makes much more sense in terms of the representation: it leaves alone
those
operators whose values are constant in a given irrep; and evolves those which
evolve the states. To turn this action explicitly into time evolution, we can
look at the action of $e^{\frac{itN}{\hbar}}$. This gives the same time
evolution as we would derive from the Heisenberg picture taking $N$ as the
Hamiltonian. For example
\eqn{heispic}{\dot{\Ad} = \frac{i}{\hbar}[N, \Ad] = \frac{i}{\hbar} \Ad  }
\eqn{heispiccomp}{\Longrightarrow \Ad (t) = e^{\frac{it}{\hbar }} \Ad
\ (= e^{\frac{itN}{\hbar }} \rhd \Ad )  }
and similarly for $A$.

Although finding a Hamiltonian for the n-body braided system might seem highly
non-trivial, the time evolution as expressed by the action {\em can} easily be
extended to n dimensions. An action on a tensor product is defined by
\cite{MajalgQFT}
\eqn{tensac}{h \rhd (v \otimes w) = h_{(1)} \rhd v \: \otimes \: h_{(2)} \rhd w
\ \ \ \ \ \ \ \ \Delta h = h_{(1)} \otimes h_{(2)}}
Since
\eqn{acton1}{Y \rhd 1 = 0 \ \ \ \ \ \ \forall Y = A, \Ad, H, N }
only one term in the coproduct of any of the generators contributes to the
action~(\ref{tensac}). This action is exactly the same as that given in
Equation~(\ref{action}), with the addition of the subscript $i$ throughout.
It is then obvious that the action of $e^{itN}$, our time evolution, is as
follows
\eqn{casac}{e^{itN} \rhd \left( \begin{array}{cc}
				\Adbs{i} & \Nbs{i} \\
				\Hbs{i}  & \Abs{i}
			\end{array} \right) = \left( \begin{array}{cc}
					e^{it} \Adbs{i} & \Nbs{i}  \\
					\Hbs{i} & e^{-it} \Abs{i}
					\end{array} \right)   }
We omit the $\hbar $ here, since it is now somewhat ambiguous as to how it
should be included, when it can take a different value for each oscillator.
It then follows, since
\Xds{i}\ is just a sum of \Ads{i}'s (and similarly for \Xs{i} ),
that the time evolution of these `diagonal generators' is
\eqn{timev}{e^{itN} \rhd \Xds{i} = e^{it} \Xds{i},\ \ \ \ \ \
e^{itN} \rhd \Xs{i} = e^{-it} \Xs{i}   }
Using these results, the time evolution of \Mbs{i} can also be calculated
\eqn{timevH}{e^{itN} \rhd \Mbs{i} = \Mbs{i} }

We can then use the isomorphism in Theorem~4.3 to find the corresponding time
evolution of the generators of \nuqh
\eqn{uqhtimev}{e^{itN} \rhd \Ads{i} = e^{it} \Ads{i},\ \ \ \ \
e^{itN} \rhd A_{i} = e^{-it} A_{i},\ \ \ \ \ e^{itN} \rhd H_{i} = H_{i},
\ \ \ \ \ e^{itN} \rhd N_{i} = N_{i}  }
It is then fairly easy, due to the relative simplicity of the unbraided
algebra,
to find an element of \nuqh\ which will act as a Hamiltonian. We need an
element
$\CH $ which has the following commutators with \Ads{i}, $A_{i}$ and $H_{i}$
\eqn{hamcomm}{[\CH, \Ads{i}] = \Ads{i},\ \ \ \ \ [\CH, A_{i}] = -A_{i},\ \ \ \
\ [\CH, H_{i}] = 0   }
The element
\eqn{Hamdef}{\CH = \sum_{i} N_{i} }
obviously meets these requirements, and is therefore a suitable Hamiltonian.
There is a very close analogy between this expression
and the usual quantum-mechanical solution for a system of non-interacting
harmonic oscillators. The action we have chosen produces the usual time
evolution  for the free system!

Having found the Hamiltonian for the unbraided algebra, it can now be mapped
back to the braided algebra via the isomorphism,
\eqn{braidham}{\bar{\CH} = \sum_{i} \Mbs{i} =
\sum_{i} \{\Nbs{i} - \omega \eHn{i}{4} \sum_{j=1}^{i-1}
(\eHn{j}{4}\,\esumHnnd{j+1}{i-1}\,\Adbs{j} \Abs{i} ) \}   }

The action of the other generators of $U_{q}(h)$ on the `diagonal generators'
of
\nb, is only slightly more complicated to calculate than that of $N$, following
a similar method. The results can then be mapped across to \nuqh\ in exactly
the same way  as was done for the time evolution. $H$ of course has a zero
action as always. The actions of the other two generators are
\begin{eqnarray}
A \rhd \left( \begin{array}{cc}
		A^{\dagger}_{i} & N_{i} \\
		H_{i} & A_{i}
	\end{array} \right)  & = & \left( \begin{array}{cc}
		\esumHnnb{1}{i-1}{2} e^{-\frac{\omega H_{i}}{4} } [H_{i}] &
\esumHnnb{1}{i-1}{2} e^{ -\frac{\omega H_{i}}{4} } A_{i}  \\
		0 & 0
				   \end{array} \right)   \nonumber  \\
 &  &    \nonumber  \\
A^{\dagger } \rhd \left( \begin{array}{cc}
		A^{\dagger }_{i} & N _{i}\\
		H_{i} & A_{i}
	\end{array} \right)  & = & \left( \begin{array}{cc}
		0 & -\esumHnnb{1}{i-1}{2} e^{ -\frac{\omega H_{i}}{4} } A^{\dagger }_{i}  \\
		0 & -\esumHnnb{1}{i-1}{2} e^{ -\frac{\omega H_{i}}{4} } [H_{i}]
				   \end{array} \right)
\end{eqnarray}

In summary, the isomorphism between the braided and unbraided tensor product
algebras reveals a hitherto unsuspected structure underlying the `normal'
unbraided system. The braided tensor product \nb\ is $U_{q}(h)$-covariant by
construction: the isomorphism demonstrates that \nuqh\ shares this property.
Furthermore, the action of $N$ on \nuqh\ which is deduced in this way turns out
to be the free particle evolution. Thus the time evolution is identified as
part of a quantum group symmetry of the n-fold harmonic oscillator. While these
results are not obvious
from the perspective of the familiar, unbraided system, we note that on the
braided side they follow automatically from standard constructions. It is to be
expected that other constructions which are quite natural from the braided
group
point of view, may similarly have important implications for the physical
system.

\end{document}